# Neural Gas Network Image Features and Segmentation for Brain Tumor Detection Using Magnetic Resonance Imaging Data

S. Muhammad Hossein Mousavi
Developer at Pars AI Company
Tehran, Iran
ORCID: 0000-0001-6906-2152
mosavi.a.i.buali@gmail.com

*Abstract*—Accurate detection of brain tumors could save lots of lives and increasing the accuracy of this binary classification even as much as a few percent has high importance. Neural Gas Networks (NGN) is a fast, unsupervised algorithm that could be used in data clustering, image pattern recognition, and image segmentation. In this research, we used the metaheuristic Firefly Algorithm (FA) for image contrast enhancement as pre-processing and NGN weights for feature extraction and segmentation of Magnetic Resonance Imaging (MRI) data on two brain tumor datasets from the Kaggle platform. Also, tumor classification is conducted by Support Vector Machine (SVM) classification algorithms and compared with a deep learning technique plus other features in train and test phases. Additionally, NGN tumor segmentation is evaluated by famous performance metrics such as Accuracy, F-measure, Jaccard, and more versus ground truth data and compared with traditional segmentation techniques. The proposed method is fast and precise in both tasks of tumor classification and segmentation compared with other methods. A classification accuracy of 95.14 % and segmentation accuracy of 0.977 is achieved by the proposed method.

## I. Introduction

The process of growing an uncommon mass of cells in the brain is called a brain tumor [1]. Brain tumor condition normally classifies into four categories of glioma tumor, meningioma tumor, not a tumor, and pituitary tumor [2] two of which (meningioma tumor and no tumor) are at the center of attention in this research. Early-stage tumor detection has high importance as it helps to prevent tumors from growing through medical approaches. Machine learning [3], image processing [4], and pattern recognition [5, 13] techniques assist this process over time and new techniques are emerging every month. Brain imaging techniques of The Electroencephalogram (EEG) [6], Positron Emission Tomography (PET) [6], and Magnetic Resonance imaging (MRI) [7] are commonly used for brain tumor detection by human experts which MRI data has more popularity among all and we are using this type of data, here. MRI devices use a strong magnetic field and radio waves to sample organs inside the body as an image [7]. Figure 1 depicts MRI dataset samples from the Kaggle platform [8, 9, 10]. Neural Gas Networks (NGN) [11] are a type of fast Artificial Neural Networks (ANN) that are used in unsupervised applications such as clustering [12], segmentation [13], and pattern recognition [5, 13]. NGN works based on feature vectors to find the most optimal data representation by expanding itself over data. By this expansion, NGN shapes itself similarly to scattered data and that's why it is a great algorithm for segmentation and clustering. NGN works based on the initial number of neurons and iterations which makes it so similar to optimization [37] and metaheuristics algorithms [14, 38]. Figure 2 illustrates 700 samples in a 2-Dimentional (2-D) space which is learned over time by NGN over 100 iterations with 50 neuron samples. Current feature extraction and segmentation techniques are not so fast and as NGN is so simple could increase the feature extraction and segmentation process significantly. Also, Figure 3 represents the structure of an ANN.

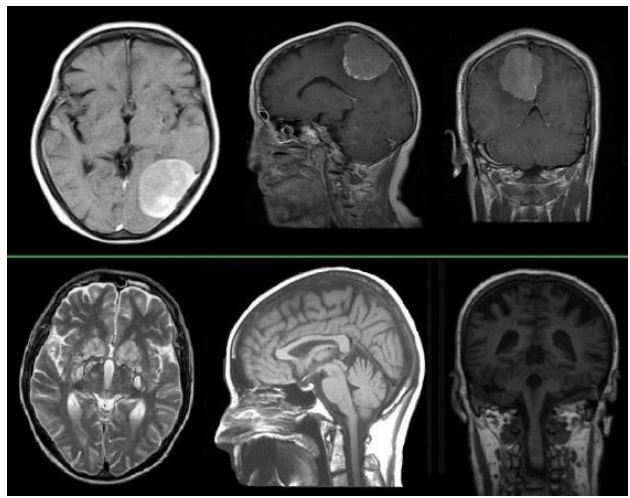

Fig. 1. MRI brain samples from the Kaggle platform (top row: meningioma tumor and bottom row: no tumor)

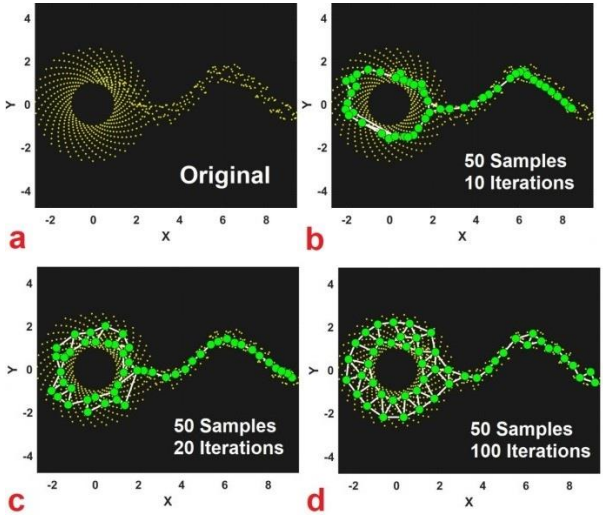

Fig. 2. a: 700 original samples (2-D space), b: learning starts with 50 neurons to mimic the original pattern (iteration 10), c: Mimicking continues till iteration 20 and, d: Mimicking and learning completes in iteration 100

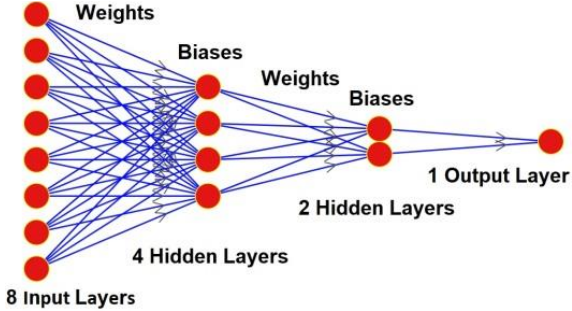

Fig. 3. The structure of an ANN with 8 input layers, 4 and 2 hidden layers, and one final output layer.

The paper consists of V main sections. The introduction describes the basics and the problem alongside our solution in general. Section II demonstrates prior related research on MRI brain tumor classification and segmentation. Section III pays to the proposed NGN features and segmentation techniques. Section IV covers evaluations and results. Finally, section V, concludes the conclusion, suggestions, and future works.

## II. LITERATURE REVIEW

As this research covers multiple subjects of feature extraction, image segmentation, and brain tumor classification, various related research to MRI brain tumor classification and segmentation are investigated and reported in Table 1. Also, due to saving space, all details related to all research are presented in Table I. These researches are conducted on different datasets in different years and some of them will be used for comparison in the evaluation and result section. According to Table I, however, some researchers used deep learning approaches, but the final accuracy doesn't represent a significant improvement over traditional methods as just two classes exist. So, it is rational to use traditional methods for this type of research as they have faster learning.

TABLE I. LITERATURE

| Authors (s) | Subject | Method | Data | Result | Date | Cite |
|---|---|---|---|---|---|---|
| Abiwinanda, et al | MRI Brain Tumor Classification | CNN | 3064 Images | 84.190 % | 2019 | [15] |
| Kharrat, et al | MRI Brain Tumor Classification | DWT-SGLDM Features Simulated Annealing Feature Selection and GA-SVM classification | 83 Images-Internet | 95.650 % | 2015 | [16] |
| Díaz-Pernas, et al | MRI Brain Tumor Segmentation and Classification | Multi-Scale CNN | 3064 Images-Internet | 97.300 % | 2021 | [17] |
| Abbood, A, et al | MRI Brain Tumor Classification | AlexNet, VGG16, GoogleNet, and RestNet50 | 3000 Images-Internet | 95.800 % | 2021 | [18] |
| Jia, Q., & Shu, H | MRI Brain Tumor Segmentation | CNN-Transformer combined model, called BiTr-Unet | BraTS202-2640 Images | Dice:0.933 | 2022 | [19] |
| Mehidi, et al | MRI Brain Tumor Segmentation | Improved K-Means | T1 Weighted-20 Images | IoU:0.870 | 2019 | [20] |
| Sharma, et al | MRI Brain Tumor Segmentation | PSO Otsu | T2 Weighted-44 Images | ACC:0.980 | 2020 | [21] |
| Ayadi, et al | MRI Brain Tumor Classification | SURF features SVM classifier | Contrast- Enhanced mri-500 | 95.830 % | 2020 | [22] |

## III. PROPOSED METHOD

The proposed method consists of two parts NGN feature extraction and NGN segmentation. For NGN features and after loading the dataset, it sends for contrast enhancement by Firefly Algorithm (FA) [23]. FA is an efficient swarm-based optimization algorithm. FA consists of five primary components: population size (fireflies), the light intensity of each firefly, coefficient of light absorption, coefficient of attraction, and mutation rate. It works by relocating fireflies with a lower light intensity to those with a higher intensity. Those fireflies with more intensity will be selected as the best candidate solutions. The full process of FA image contrast enhancement is fully expressed in Figure 4. As mentioned earlier, NGN expands and mimics itself as connected graphs over pattern data and over iterations which is proper for segmentation, clustering, and feature extraction applications.

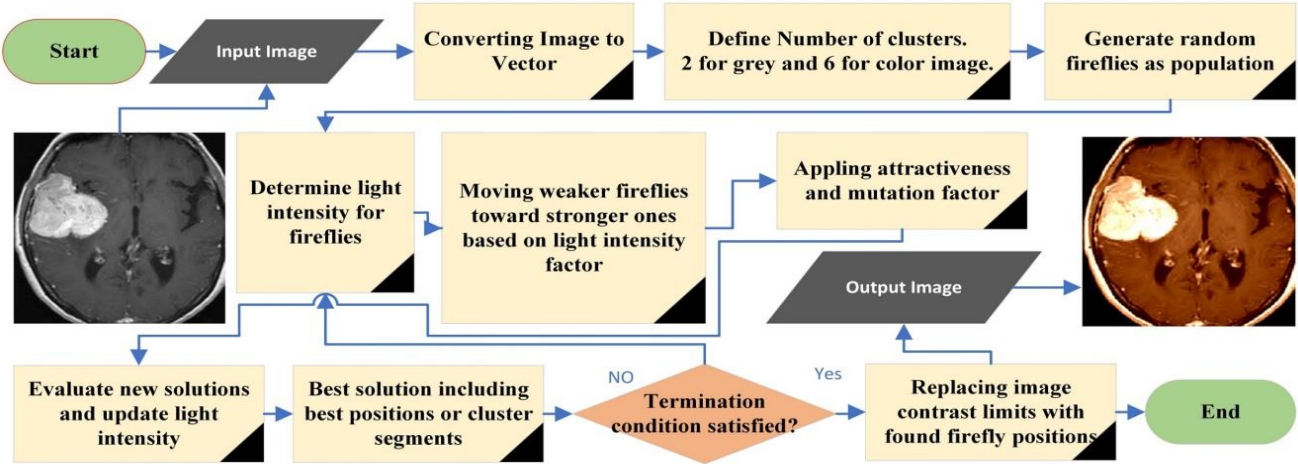

Fig. 4. FA image contrast enchantment flowchart

The steps of NGN [11] are as follows. Having a probability distribution $P(x)$ of vectors $x$ and a number of feature vectors $w_i\ i = 1, \cdots, N$. In each step $t$, a vector $x$ randomly selected from $P(x)$ is showed. Later on, the distance order of the feature data vectors to the given data vector will be determined. Let $i_0$ be the closest feature vector index, $i_1$ the index of the second closest feature vector, and $i_{N-1}$ the index of the feature vector most distant to $x$. Then each feature vector is adapted according to (1).

$$NGN\ Model = w_{i_k}^{t+1} = w_{i_k}^t + \varepsilon \cdot e^{-k/\lambda} \cdot (x - w_{i_k}^t), k = 0, \cdots, N-1 \quad (1)$$

With ε as the adaptation step size and λ as the neighborhood range. ε and λ getting reduced with increasing t. After enough adaptation steps, the feature vectors cover the data space with the minimum error possible.

The NGN feature extraction part starts with contrast enhancement by FA and converting the image matrix to a vector for further processing. After defining the number of neurons and iterations for NGN, the learning process initiates. Weights are extracted and summed. Finally, Lasso regularization [24] applies summed weights for feature selection or dimensionality reduction step and sends labeled data for classification. Figure 5 depicts the proposed NGN feature extraction method as a flowchart. In this figure, each image dimension is considered 256*256. NGN segmentation is similar to NGN feature extraction but with weight summation, cell values round and reshape into 1 and number of segments. Then, the vector converts back into an indexed image matrix. Finally, a color assigns to each index which shapes the segmented image. Figure 6 represents the flowchart of the proposed NGN image segmentation method.

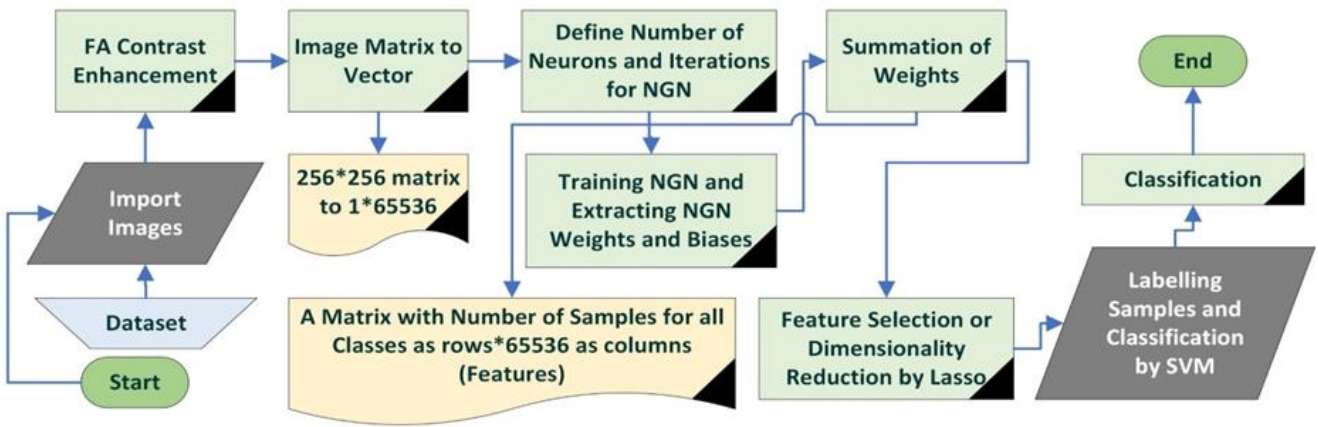

Fig. 5. Flowchart of the proposed NGN feature extraction method

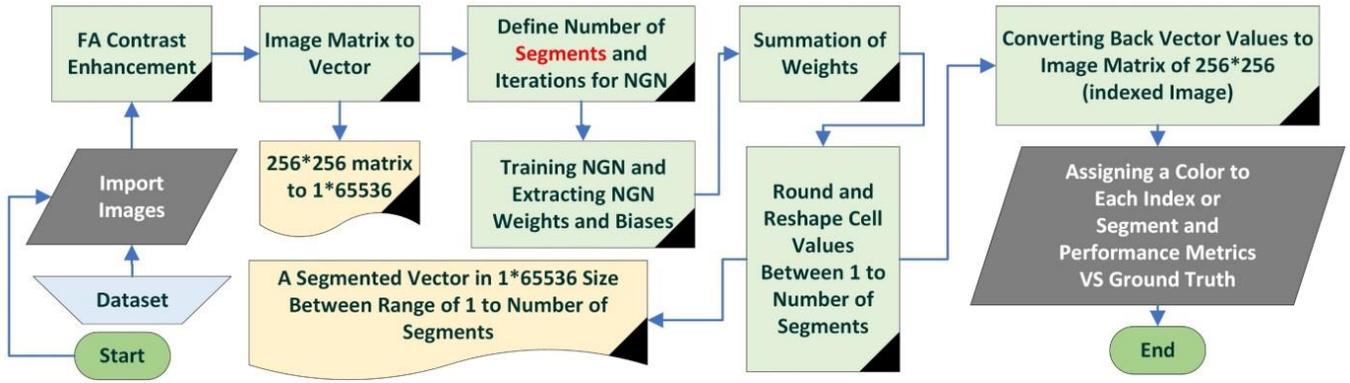

Fig. 6. Flowchart of the proposed NGN image segmentation method

IV. EVALUATIONS AND RESULTS

In order to validate the proposed segmentation method's final result, the predicted output should be transformed into a Black and White (BW) image in order to compare with the mask or Ground Truth (GT) BW image. Also, there are definitions such as Positive (P), Negative (N), True Positive (TP), True Negative (TN), False Positive (FP), and False Negative (FP) [25] which are needed for evaluation. P is the real number of positive cases and N is the real number of negative cases in the image. By considering white color or 1 as positive and black color or 0 as negative in the BW image, TP is when the model correctly predicts the positive class and TN is when the model correctly predicts the negative class. FP is when the model incorrectly predicts the positive class and FN is when the model incorrectly predicts the negative class [25, 26]. Figure 7 shows mentioned concept as a visual. Table II shows FA and NGN parameters that are used in this experiment. Three datasets [8, 9, 10] from Kaggle are employed in this research which converted into two. Dataset one consists of 200 samples from [8] and [10] and dataset 2 is consisted of 400 selected samples from [9]. Vividly K-means [27] algorithm is able to cluster close groups of samples into the various number of clusters based on their distance from each other in the space. The same technique could be used for image pixel values which, could be gray images or color. Another segmentation method is the Watershed image segmentation algorithm [28]. This is a region-based algorithm based on drainage in basins or rivers. It works based on the water level as water height presents the separating line from one region or segment to another. A popular segmentation technique is called Otsu's thresholding segmentation [29] which is categorized as a threshold-based segmentation technique. It works as a thresholding process for pixel intensity values in order to minimize interclass variance for separating pixels into two classes foreground and background. The number of the threshold value determines, the number of segments. Figure 8 illustrates, Comparison of results on an MRI brain tumor sample by different algorithms.

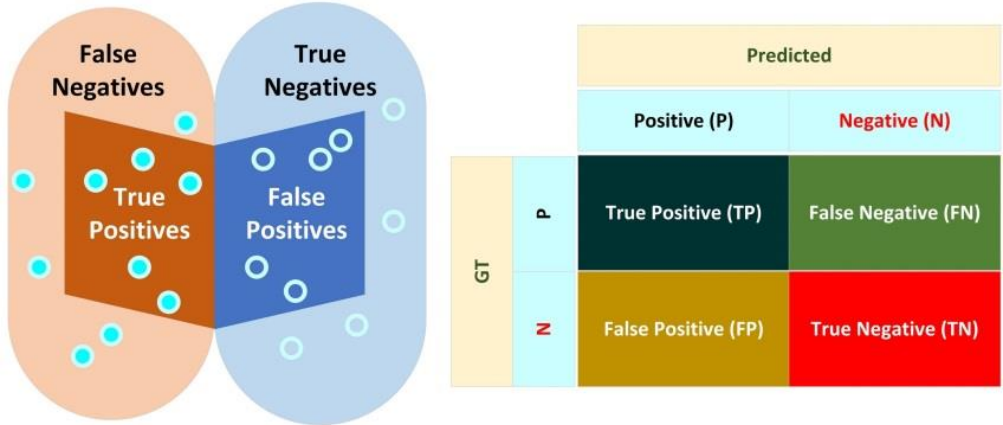

Fig. 7. Ground Truth, Predicted, TP, FN, FP, and TN as a visual

Now, Accuracy [25, 26] is the percent of pixels that are correctly classified as (2):

$$Accuracy = \frac{TP+TN}{TP+TN+FP+FN} \qquad (2)$$

Precision [25, 26] means how many of those predicted objects had matching ground truth annotation and is calculated as (3):

$$Precision = \frac{TP}{TP+FP} \qquad (3)$$

Recall [25, 26] or sensitivity means of all objected annotated in the ground truth, how many are captured as positive predicator and is calculated a (4):

$$Recall = \frac{TP}{TP+FN} \qquad (4)$$

The boundary F1 (BF) [25, 26] contour matching score or f-measure determines how decent the predicted boundary of each class aligns with the true boundary in the image. The BF score is defined as the harmonic mean (F1-measure) of the precision and recall values with a distance error tolerance and defined as (5):

$$F1 - Measure = \frac{2*TP}{2*TP+FP+FN} \quad (5)$$

Intersection over Union (IoU) [25, 26] also, known as the Jaccard similarity coefficient, is one of the most segmentation metrics. IoU is a statistical accuracy measurement that penalizes false positives. For each class, IoU is the ratio of correctly classified pixels to the total number of ground truth and predicted pixels in that class based on (6):

$$IoU = \frac{F1-Measure}{2-F1-Measure} \quad (6)$$

Tables III and IV present returned results for comparing the proposed NGN segmentation method with Otsu, Watershed, and K-Means segmentation methods for Accuracy, Precision, Recall, F-Measure, and IoU performance metrics plus runtime in seconds for Datasets 1 and 2, respectively. Clearly Proposed NGN segmentation method outperformed other methods versus ground truth data. Also, K-means and Otsu placed in second and third places. Watershed segmentation had poor performance compared with others. Also, Watershed is the fastest algorithm, K-means is the slowest, and the proposed NGN is in second place regarding runtime speed.

There is a type of feature that is based on the edge, place, and angle of the pixels. It is possible to extract these features using image gradients. They are Histograms of Oriented Gradient (HOG) [30] features. These features are local-based. In edge-based features which are possible to get by the gradient of the image, useful information is extracted from angles and position of the connected pixels. HOG features are in horizontal, vertical, and diagonal directions. HOG features are extracted from blocks with different sizes. These blocks have two values of magnitude and direction. Magnitude determines the scale of the block and direction determines the path which that specific edge follows. Local Phase Quantization (LPQ) [31] is a frequency neighborhood-based feature based on Fourier transform [32]. It manipulated the blurring effect in magnitude and phase channels. Phase channel is capable of deactivating low pass filters that exist in some images. Gabor filters [33] are linear filters for texture analysis which could be described as features. These features are not sensitive to rotation, resizing, and illumination changes. They are based on texture and robust against low pass filters which are nice for edge detection. If they are adjusted very well, they could have very precise performance. They have great responses to sudden changes which makes these filters very good in MRI data analysis. Their main advantages are a change in illumination, rotation, and resizing. Speeded Up Robust Features (SURF) [34] are feature detectors and descriptor algorithms. SURF is so fast algorithm and has great resistance against rotation. First, the image integral calculates; Then, feature points using the hessian algorithm [35] will be found. Making scale space is the third step. Determining the maximum point is the next step. Finally, the feature vector will be made by repeating the preview steps.

TABLE II. FA AND NGN PARAMETERS USED FOR THIS RESEARCH

| Parameters | FA | NGN |
|---|---|---|
| Decision Variables (DV) | Number of Clusters | - |
| Iterations | 20 | 20 |
| Population Size (P) | 10 | - |
| Mutation Rate | 0.2 | - |
| Light Absorption Coefficient | 1 | - |
| Attraction Coefficient | 2 | - |
| Mutation Damping Ratio | 0.98 | - |
| Number of Neurons | - | 5 |
| Epsilon Initial | - | 0.3 |
| Epsilon Final | - | 0.02 |
| Lambda Initial | - | 2 |
| Lambda Final | - | 0.1 |
| T Initial | - | 5 |
| T Final | - | 10 |

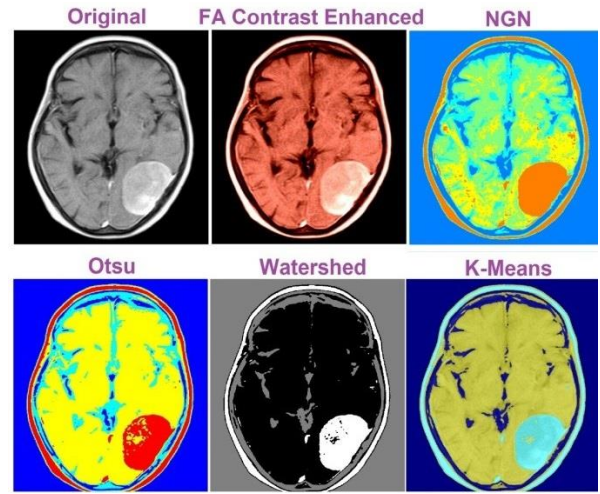

Fig. 8. Segmentation Results Comparison (four segments)

Tables V and VI represent classification accuracy by the SVM classifier on datasets 1 and 2 in a comparable manner, respectively. Data is divided into 70 % training and 30 % testing for the classification task. Higher accuracy for dataset 1 goes to NGN features and for dataset 2 goes to the CNN method. Clearly, more data in deep learning methods provide better results. As dataset 1 had just 200 samples CNN could not achieve the best result but by doubling the number of samples in dataset 2 to 400 samples, CNN outperformed others. Other than that the proposed NGN features returned very decent results compared with other features both in the train and test phases. Additionally, Figure 9 depicts the training confusion matrix for NGN features by SVM classifier alongside its related Receiver Operating Characteristic (ROC) curve [36] for both datasets. Table VII contains the runtime for feature extraction methods in seconds. As it mentioned earlier, one of the advantages of NGN is its high speed. Clearly, it is the fastest feature extraction method in this table.

TABLE III. SEGMENTATION RESULTS FOR DATASET 1 (200 SAMPLES VERSUS GROUND TRUTH) AND RUNTIME IN SECONDS

| Dataset 1 | Accuracy | Precision | Recall | F-Measure | IoU | Runtime |
|---|---|---|---|---|---|---|
| Otsu | 0.885 | 0.899 | 0.866 | 0.880 | 0.815 | 1.414 |
| Watershed | 0.814 | 0.857 | 0.807 | 0.831 | 0.788 | **1.009** |
| K-Means | 0.918 | 0.923 | 0.910 | 0.922 | 0.871 | 5.904 |
| **NGN** | **0.944** | **0.950** | **0.938** | **0.969** | **0.903** | 1.187 |

TABLE IV. SEGMENTATION RESULTS FOR DATASET 2 (400 SAMPLES VERSUS GROUND TRUTH) AND RUNTIME IN SECONDS

| Dataset 2 | Accuracy | Precision | Recall | F-Measure | IoU | Runtime |
|---|---|---|---|---|---|---|
| Otsu | 0.936 | 0.955 | 0.918 | 0.930 | 0.890 | 2.914 |
| Watershed | 0.921 | 0.940 | 0.911 | 0.919 | 0.881 | **2.211** |
| K-Means | 0.949 | 0.965 | 0.938 | 0.934 | 0.920 | 11.008 |
| **NGN** | **0.977** | **0.989** | **0.951** | **0.965** | **0.949** | 2.406 |

TABLE V. SVM CLASSIFICATION ACCURACY FOR DATASET 1

| Dataset 1 | SURF | LPQ | HOG | Gabor-Filters | CNN | NGN |
|---|---|---|---|---|---|---|
| Train | 96.24 | 95.64 | 94.51 | 95.55 | 97.20 | **97.66** |
| Test | 94.80 | 93.17 | 92.30 | 91.40 | 94.29 | **95.14** |

TABLE VI. SVM CLASSIFICATION ACCURACY FOR DATASET 2

| Dataset 2 | SURF | LPQ | HOG | Gabor-Filters | CNN | NGN |
|---|---|---|---|---|---|---|
| Train | 95.71 | 94.70 | 92.25 | 94.87 | **98.31** | 95.99 |
| Test | 92.66 | 93.22 | 90.88 | 90.39 | **94.60** | 94.10 |

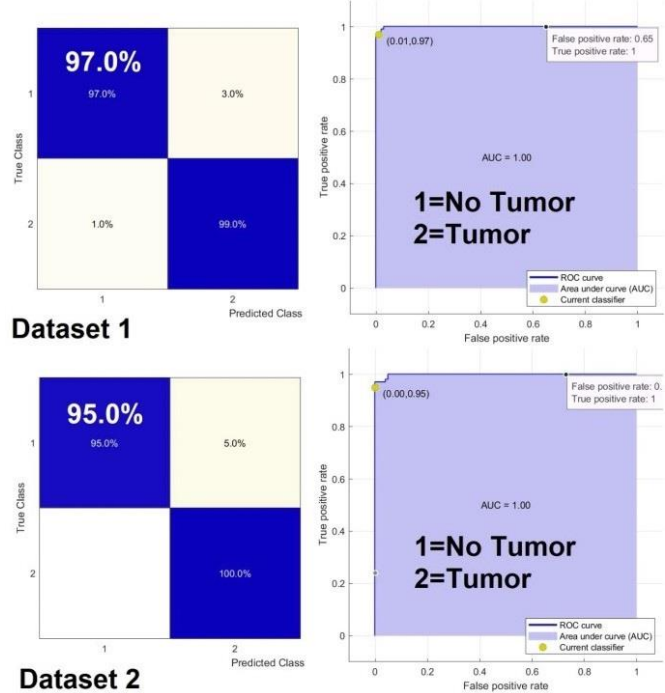

Fig. 9. Training confusion matrix for NGN features by SVM classifier alongside its related ROC curve for both datasets.

TABLE VII. FEATURE EXTRACTION RUNTIME IN SECONDS (S) - (CORE I-7 CPU, 16 GB OF RAM, SSD HARD DRIVE)

| Runtime | SURF | LPQ | HOG | Gabor-Filters | CNN | NGN |
|---|---|---|---|---|---|---|
| Dataset 1 | 6.744 | 3.724 | 0.769 | 5.120 | 20.550 | **0.279** |
| Dataset 2 | 12.994 | 7.219 | 1.510 | 9.997 | 41.607 | **0.498** |

## V. CONCLUSION, SUGGESTIONS, AND FUTURE WORKS

By employing NGN, it is possible to extract decent features out of signals and images for classification in a low number of classes at a very high speed. As using NGN returns highly precise segmentation results in a fraction of time and has low complexity, it could be used in low-power processing units such as robots or minicomputers in real-time. NGN features and segmentation returned robust performance compared with traditional and modern deep learning techniques but with less computational complexity. Testing NGN features for object detection was a side experiment and could classify 8 different small home objects with 91.00 % accuracy in the test phase. It is suggested to use Growing NGN (GNG) for segmentation and feature extraction tasks as it might return even better results as GNG is an improved version of NGN. Using NGN for data clustering and making a real-time NGN segmentation system with a webcam is of future work. Also, improving weights and biases of trained NGN by optimization algorithms such as FA is another suggestion for the final reader.